# A Capsule-Sized Multi-Wavelength Wireless Optical System for Edge-AI-Based Classification of Gastrointestinal Bleeding Flow Rate

Yunhao Bian[1], Dawei Wang[1]*, Mingyang Shen[1], Xinze Li[1], Jiayi Shi[1], Ziyao Zhou[1], Tiancheng Cao[1], and Hen-Wei Huang[1,2]*

*Abstract*— Post-endoscopic gastrointestinal (GI) rebleeding frequently occurs within the first 72 hours after therapeutic hemostasis and remains a major cause of early morbidity and mortality. Existing non-invasive monitoring approaches primarily provide binary blood detection and lack quantitative assessment of bleeding severity or flow dynamics, limiting their ability to support timely clinical decision-making during this high-risk period. In this work, we developed a capsule-sized, multi-wavelength optical sensing wireless platform for order-of-magnitude–level classification of GI bleeding flow rate, leveraging transmission spectroscopy and low-power edge artificial intelligence. The system performs time-resolved, multi-spectral measurements and employs a lightweight two-dimensional convolutional neural network for on-device flow-rate classification, with physics-based validation confirming consistency with wavelength-dependent hemoglobin absorption behavior. In controlled in vitro experiments under simulated gastric conditions, the proposed approach achieved an overall classification accuracy of 98.75% across multiple bleeding flow-rate levels while robustly distinguishing diverse non-blood gastrointestinal interference. By performing embedded inference directly on the capsule electronics, the system reduced overall energy consumption by approximately 88% compared with continuous wireless transmission of raw data, making prolonged, battery-powered operation feasible. Extending capsule-based diagnostics beyond binary blood detection toward continuous, site-specific assessment of bleeding severity, this platform has the potential to support earlier identification of clinically significant rebleeding and inform timely re-intervention during post-endoscopic surveillance.

## I. Introduction

Post-endoscopic gastrointestinal (GI) rebleeding remains a major clinical challenge and a leading cause of early morbidity following therapeutic endoscopy. Acute GI bleeding affects an estimated 80–150 per 100,000 adults annually [1], and despite successful initial hemostasis, reported early rebleeding rates range from approximately 8% to 25% within the first 72 hours after intervention, depending on lesion characteristics and clinical setting [2], [3]. This early post-procedural period represents the highest-risk window for clinical deterioration, often requiring urgent repeat endoscopy, blood transfusion, or surgical escalation, and is associated with reported in-hospital mortality rates of approximately 5%–14% [2]–[4]. However, current post-endoscopic surveillance relies primarily on intermittent clinical indicators such as hemoglobin decline and hemodynamic instability, which are inherently delayed and may fail to detect early or low-grade bleeding. These limitations motivate the need for technologies that enable continuous, site-specific monitoring of post-endoscopic bleeding dynamics.

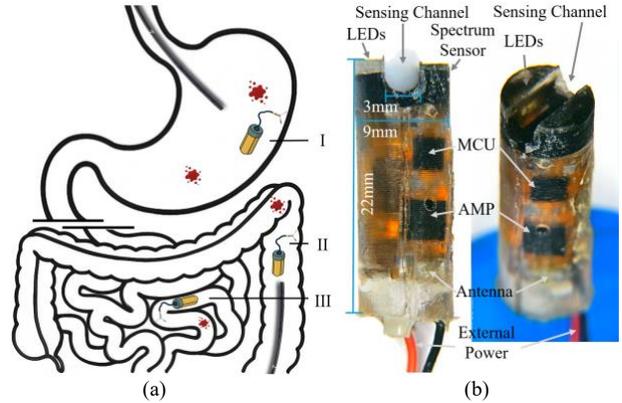

Fig. 1. Overview of the proposed capsule-based system for site-specific gastrointestinal bleeding monitoring. (a) Schematic of capsule deployment using standard endoscopic clip fixation in different gastrointestinal regions: (I) stomach, (II) colon, and (III) small intestine. (b) Photograph of the capsule prototype used in this study.

Upper GI endoscopy is the first-line diagnostic and therapeutic modality for nonvariceal upper GI bleeding [2]. Bleeding severity assessment is predominantly qualitative, relying on visual scoring systems such as the Forrest classification, and is highly dependent on intragastric conditions. Blood clots, food residue, bubbles, and foam frequently degrade visualization quality, resulting in intermittent and subjective assessment without continuous quantification of bleeding rate. When endoscopic findings are inconclusive or bleeding is intermittent, scintigraphy and cross-sectional imaging are used as complementary diagnostic tools [4]. Radionuclide imaging using technetium-99m–labeled red blood cells can detect bleeding rates as low as approximately 0.1 mL/min [5], while computed tomography angiography (CTA) provides rapid localization with detection thresholds of approximately 0.3–0.5 mL/min [6], [7]. However, these modalities are invasive, non-continuous,

1. School of Electrical and Electronic Engineering, Nanyang Technological University, Singapore.
2. Lee Kong Chian School of Medicine, Nanyang Technological University, Singapore.

This work is supported by the A*STAR MTC Programmatic Seed Fund (M24N9b0130) and the Nanyang Assistant Professorship.

*Corresponding authors: Dawei Wang (email: dawei.wang@ntu.edu.sg), Hen-wei Huang (email: henwei.huang@ntu.edu.sg).

resource-intensive, and unsuitable for bedside or longitudinal monitoring.

Capsule endoscopy has become an established diagnostic modality for GI bleeding and small-bowel disorders, as reflected in recent clinical guidelines [8]. Building upon this foundation, artificial intelligence has been increasingly applied to capsule endoscopy for automated detection of bleeding events using image-based analysis, demonstrating improved diagnostic efficiency and reduced clinician workload [9]. In parallel, non-imaging ingestible sensing approaches have been developed to directly detect intraluminal blood without reliance on visualization. Early telemetric capsule systems such as HemoPill® have demonstrated the feasibility of real-time blood detection within the GI tract [10]. More recent optical and electrochemical ingestible sensors further extend this paradigm by enabling continuous intraluminal monitoring under controlled experimental or preclinical conditions [11], [12]. Despite these advances, both imaging- and non-imaging-based capsule systems remain largely limited to binary or coarse ordinal blood detection and typically lack stable localization at high-risk bleeding sites. Consequently, they are unable to provide sustained, site-specific, or quantitative assessment of bleeding dynamics—such as flow-rate classification—during the critical post-intervention period.

In this work, we present a capsule-sized, multi-wavelength optical sensing platform and an experimental validation framework for order-of-magnitude–level classification of gastrointestinal bleeding flow rate. The device is designed for site-specific deployment using standard endoscopic clips, enabling localized and continuous monitoring at different regions of the gastrointestinal tract (Fig. 1(a)), while the capsule prototype used in this study is shown in Fig. 1(b). Among these scenarios, the gastric environment represents a particularly challenging and clinically relevant worst case due to its complex contents, dynamic mixing, and variable acidity. Accordingly, the present study focuses on controlled in vitro experiments under simulated gastric conditions to evaluate the feasibility of extending capsule-based diagnostics from binary blood detection toward quantitative, clinically actionable assessment of bleeding severity. To support continuous monitoring under the stringent power and bandwidth constraints of ingestible devices, edge AI processing is further integrated directly on the capsule electronics, enabling on-device flow-rate classification and substantially reducing the need for high-volume wireless data transmission.

## II. METHODS

This section describes the proposed capsule-sized multi-wavelength optical sensing system, the underlying measurement principles, and the data-driven model used for flow-rate state classification. Specifically, we first introduce the system architecture and in vitro experimental setup, followed by a physics-based analysis of multi-wavelength optical measurements, and finally present the time–spectral representation and the lightweight 2D convolutional neural network employed for classification.

### A. System Architecture and In Vitro Experimental Setup

The proposed capsule-sized optical sensing system and in vitro experimental setup are illustrated in Fig. 2. The system acquires time-resolved, multi-spectral transmission measurements under controlled experimental conditions. As shown in Fig. 2(a), the capsule integrates multiple light-emitting diodes (violet 405 nm, green 515–555 nm, near-infrared 850 nm, and white, 6000 K), a spectral sensor, and a microcontroller unit (MCU, STM32WBA65). Transmitted light is captured by a compact spectral sensor (AS7343) providing 12 discrete channels spanning the visible and near-infrared range, with center wavelengths at 405, 425, 450, 475, 515, 550, 555, 600, 640, 690, 745, and 855 nm. During each measurement cycle, two sequential illumination states are used to acquire a 24-dimensional spectral vector (ch1–ch24). For clarity, although the AS7343 sensor provides only 12 physical spectral channels, the two consecutive 12-channel measurements acquired under different illumination states are concatenated to form a 24-dimensional measurement vector. Specifically, the violet, green, and near-infrared LEDs are first activated to record ch1–ch12, followed by activation of the white LED to record ch13–ch24. Hereafter, the term "channel" is used to refer to the elements of this 24-dimensional vector for notational convenience.

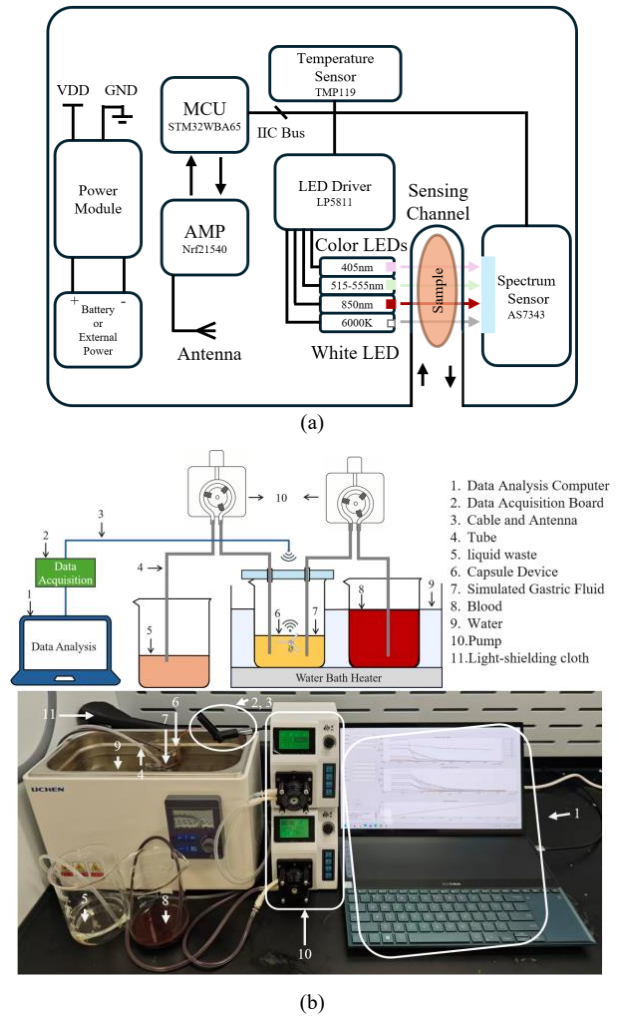

Fig. 2. Experimental system. (a) Capsule prototype system overview, and (b) in vitro experimental setup.

The capsule prototype measures 22 mm in length and 9 mm in diameter, with an optical sensing channel width of 3 mm, and is compatible with a standard size-000 capsule shell. To facilitate the experiment, the capsule was powered via cable instead of using a battery. Data were wirelessly collected via Bluetooth Low Energy (BLE) using an external USB Bluetooth adapter connected to a host computer. The in vitro experimental setup is shown in Fig. 2(b). The capsule sensor is placed in a beaker containing the test medium and positioned within a temperature-controlled water bath at 37 ºC. During the experiments, the water bath was covered with a light-shielding cloth to minimize interference from ambient light. A magnetic stirrer ensures homogeneous mixing, while sterile anticoagulated porcine blood is infused at controlled rates using a programmable syringe pump to simulate gastrointestinal bleeding. The infusion rate is defined as the equivalent bleeding flow rate (mL/min). Flow-rate levels are selected to span clinically relevant bleeding regimes based on reported detection limits of established gastrointestinal bleeding imaging modalities, ranging from approximately 0.1 mL/min to 1.0 mL/min [4]. Accordingly, six classes including one non-bleeding interference class (0.0 mL/min) and five bleeding flow-rate levels ranging from 0.1 to 0.9 mL/min are used in this study. Deionized water and simulated gastric fluid (SGF) were used as background media. Nonzero bleeding conditions are simulated by introducing blood into SGF, while the no-bleeding condition (0.0 mL/min) consists of independent experiments in water containing single non-blood interference substances. To represent the diversity of non-blood intragastric contents encountered in clinical scenarios, a set of visually and spectrally distinct interference mixtures was selected, including beverages (e.g., tea, coffee, grape juice, and cola), fermented drinks (e.g., kvass), dairy-based liquids (e.g., fresh milk and chocolate milk), vegetable-based fluids (e.g., beetroot and tomato sauce), and acidic juices (e.g., grapefruit juice). Each interference mixture was tested independently under identical experimental conditions and aggregated into a single non-bleeding interference class (0.0 mL/min) during model training and evaluation.

### B. Measurement Principle

The proposed system performs transmission-based optical measurements at multiple wavelengths. For a given wavelength $\lambda$, the sensor output is proportional to the optical intensity transmitted through the medium. Assuming that variations in optical path length and scattering remain relatively stable over short time intervals, the transmission behavior can be approximated by the Beer–Lambert law:

$$I(\lambda) = I_0(\lambda) exp\ (-\mu(\lambda)cL) \quad (1)$$

where $I_0(\lambda)$ denotes the incident intensity, $\mu(\lambda)$ is the wavelength-dependent absorption coefficient, $c$ represents the effective hemoglobin concentration, and $L$ is the effective optical path length.

Before introducing data-driven classification, a rapid physics-based prior analysis is conducted to verify that the measured signals exhibit behavior consistent with the known optical properties of hemoglobin. To this end, a ratio of transmitted intensities measured at two representative wavelengths is constructed. Specifically, $\lambda_1$ is selected from a relatively strong absorption region, while $\lambda_2$ is selected from a weaker absorption region. The ratio of transmitted intensities can be expressed as:

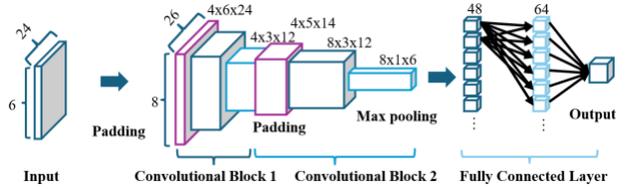

Fig. 3. Architecture of the proposed Tiny 2D-CNN for bleeding flow-rate classification.

$$\frac{I(\lambda_1)}{I(\lambda_2)} = \frac{I_0(\lambda_1)}{I_0(\lambda_2)} exp\ (-[\mu(\lambda_1) - \mu(\lambda_2)]cL) \quad (2)$$

For convenience, this relationship can be rewritten as:

$$\frac{I(\lambda_1)}{I(\lambda_2)} = C \cdot exp\ (-\Delta\mu cL) \quad (3)$$

where $C = \frac{I_0(\lambda_1)}{I_0(\lambda_2)}$ and $\Delta\mu = \mu(\lambda_1) - \mu(\lambda_2)$.

If the measured intensity ratio at different wavelengths follows the expected exponential dependence, this behavior directly indicates that the sensing system operates in a manner consistent with the underlying optical absorption physics of hemoglobin. Accordingly, the ratio-based formulation is used solely as a rapid physical validation of the measurement behavior, rather than for quantitative estimation or as an input feature for classification.

### C. Tiny 2D-CNN Architecture for Flow-Rate Classification

The extracted time–spectral measurements are classified using a lightweight two-dimensional convolutional neural network (Tiny 2D-CNN), whose architecture is illustrated in Fig. 3. The network is specifically designed to balance classification performance and computational efficiency for embedded deployment. The input to the network is a two-dimensional time–spectral representation with a size of 6×24, where the temporal dimension corresponds to six consecutive time steps and the spectral dimension corresponds to the 24-dimensional measurement vector acquired within each sensing cycle. Zero padding is applied prior to convolution to preserve spatial resolution along both dimensions. The first convolutional block consists of a 3×3 convolution with 4 output channels, followed by a rectified linear unit (ReLU) activation and max pooling, producing feature maps of size 4×3×12. The second convolutional block applies an additional 3×3 convolution to increase the channel depth to 8, followed by ReLU activation and max pooling, resulting in compact feature maps of size 8×1×6. The resulting feature maps are flattened into a 48-dimensional feature vector, which is passed to a fully connected layer with 64 hidden units. The final output layer produces the flow-rate classification result. This compact architecture limits both memory footprint and computational complexity, enabling efficient on-device inference while preserving discriminative capability for bleeding flow-rate classification.

## III. RESULTS AND DISCUSSION

This section presents the experimental results of the proposed system, including physics-based validation of the optical measurements, data-driven bleeding flow-rate classification, and on-device deployment with energy consumption evaluation.

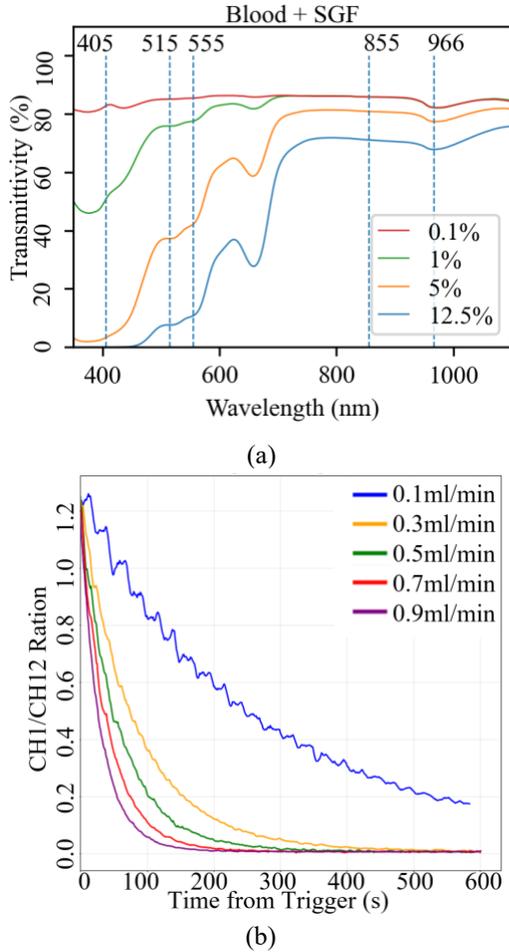

Fig. 4. Wavelength-dependent optical characterization of blood. (a) Transmission spectra. (b) Temporal evolution of the ch1/ch12 ratio.

*A. Physical Validation*

To verify the physical consistency of the proposed sensing system, a set of physics-based validation experiments was conducted. Fig. 4(a) presents the hemoglobin absorption spectrum measured using a benchtop spectrophotometer (SHIMADZU UV-1800), confirming the expected wavelength-dependent absorption behavior across the visible and near-infrared range. This measurement serves as an external physical reference for the multi-wavelength transmission sensing performed by the capsule system, establishing that the optical bands selected in this study are well aligned with known hemoglobin absorption characteristics.

Based on this reference, Fig. 4(b) shows the ratio of transmitted intensities measured at ch1 (405 nm) and ch12 (855 nm), corresponding to representative strong- and weak-absorption regions, respectively. Both channels are acquired within the same illumination state under fixed LED excitation conditions, such that the constructed ratio reflects wavelength-dependent absorption differences rather than illumination-state variations. As the bleeding flow rate increases, the measured intensity ratio exhibits a clear monotonic decrease, consistent with the exponential dependence predicted by the Beer–Lambert-based analysis. This behavior indicates that the capsule-based measurements are dominated by hemoglobin absorption rather than system-specific artifacts, and that the proposed optical sensing system operates in a manner consistent with the underlying physical model. The ratio-based analysis is therefore used solely as a rapid physical sanity check, rather than as an input feature or estimator for subsequent classification.

In addition, supplementary experiments were conducted to examine the influence of gastric acidity on the measured spectral behavior. Specifically, identical blood volumes and flow rates were tested in deionized water and simulated gastric fluid (SGF, pH ≈ 1.5). To enable comparison across experimental conditions, the resulting 24-channel time–spectral measurements were standardized using Z-score normalization, which is commonly adopted in data-driven analysis to ensure consistent scaling across channels and experiments. Under identical flow-rate conditions, the normalized spectral responses from the two media exhibited highly consistent channel-wise temporal profiles and relative spectral shapes. This observation indicates that gastric acidity does not introduce systematic distortion in the hemoglobin-dominated multi-wavelength measurements captured by the system. As a result, physiologically relevant pH variability is unlikely to confound the learned spectral–temporal representations, supporting the feasibility of flow-rate level classification using multi-channel time–spectral measurements in dynamic gastric environments.

*B. Bleeding Flow-Rate Classification Performance*

The classification performance of the proposed Tiny 2D-CNN is evaluated using a six-class test set comprising one non-bleeding interference class (0.0 mL/min) and five bleeding flow-rate levels (0.1–0.9 mL/min). Fig. 5(a) shows the resulting confusion matrix, achieving an overall classification accuracy of 98.75%. The interference class, corresponding to the no-bleeding condition (0.0 mL/min), is correctly identified in all test cases, with no misclassification into bleeding-related categories. This result indicates that the network effectively distinguishes blood-related optical dynamics from non-blood gastrointestinal interference. For bleeding conditions, classification errors are limited to adjacent flow-rate levels, with no cross-level misclassification observed. This behavior reflects the gradual and continuous nature of bleeding dynamics, where neighboring flow-rate categories exhibit partially overlapping temporal–spectral characteristics.

To further examine the structure of the learned feature representations beyond classification accuracy, Fig. 5(b) presents a t-SNE visualization of feature embeddings extracted from the final hidden layer of the network. In contrast to Fig. 5(a), where all non-bleeding samples are aggregated into a single interference class, the interference samples in Fig. 5(b) are decomposed by mixture type for visualization purposes. Distinct and compact clusters are observed for most non-blood interference mixtures, indicating that the learned feature space preserves meaningful intra-class structure among heterogeneous intragastric contents, even though these samples are treated as a single class during training and evaluation.

Notably, certain interference mixtures (e.g., tea- and fermented drink–based samples) appear closer to blood-containing samples in the embedded feature space, while others (e.g., beetroot juice, dairy-based mixtures, and coffee) remain well separated. This relative organization is consistent with varying degrees of spectral–temporal similarity between different non-blood contents and hemoglobin-dominated signatures. Overall, the t-SNE analysis suggests that the proposed 24-channel time–spectral representation and data-driven feature learning capture discriminative patterns beyond a simple binary separation, supporting robust bleeding flow-rate classification under complex and diverse gastrointestinal interference conditions.

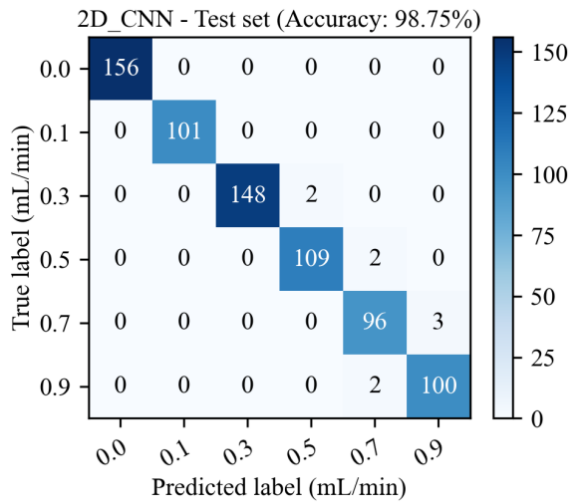

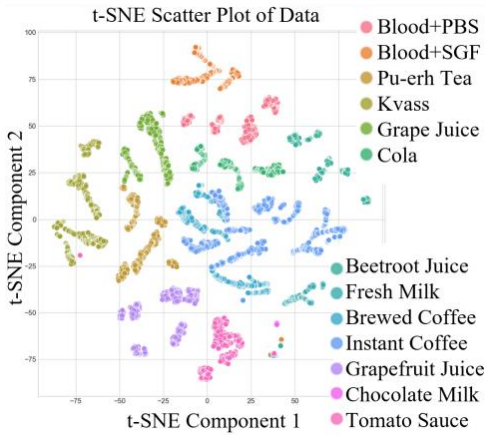

Fig. 5. Bleeding flow-rate classification results. (a) Confusion matrix. (b) t-SNE visualization.

TABLE I. POWER CONSUMPTION UNDER DIFFERENT CONFIGURATIONS

| Case | Active Time | MCU Energy | Low-Power Mode | Standby Current (µA) |
|---|---|---|---|---|
| 1 | 8.16ms | 16.45nAh | Standby | 4.5 |
| 2 | 4.2s | 5.165µAh | Standby | 4.5 |
| 3 | 4.6ms | 9.73nAh | Stop | 65.3 |

| Case | Standby Time (s) | Standby Energy (µAh) | Total Energy (µAh) | Total Charge (µC) |
|---|---|---|---|---|
| 1 | 59.99 | 0.07499 | 0.09144 | 329.17 |
| 2 | 59.99 | 0.06975 | 5.13515 | 18486.53 |
| 3 | 60 | 1.08825 | 1.09798 | 3952.73 |

### C. Embedded Inference and Power Consumption

The proposed Tiny 2D-CNN was deployed on the STM32WBA65 microcontroller using the STM32 X-CUBE-AI toolchain to evaluate on-device inference efficiency and power consumption. Measurements were conducted over a one-minute duty cycle consisting of system startup, inference or data transmission, and low-power idle operation. The quantitative results are summarized in Table I.

For inference-based operation, the MCU active energy consumption remains below 0.02 µAh per cycle (Cases 1 and 3), indicating that the computational cost of the proposed 2D-CNN is negligible compared to the overall system energy budget. When the system operates in standby mode following inference (Case 1), the low quiescent current of approximately 4.5 µA results in a total energy consumption of only 0.091 µAh per minute.

In contrast, direct data transmission without on-device inference (Case 2) incurs substantially higher energy consumption. Continuous MCU operation for Bluetooth communication results in a total energy usage of 5.14 µAh per minute, which is more than 50× higher than the inference-based standby configuration. Although stop mode further reduces MCU active energy (Case 3), the higher stop-mode current (65.3 µA) leads to increased standby energy, yielding a total consumption of approximately 1.10 µAh per minute.

From a system-level perspective, performing on-device inference prior to wireless transmission provides significant energy savings. Under a representative operating scenario, performing nine local inference cycles followed by a single data transmission consumes approximately 5.96 µAh, compared to 51.35 µAh required for transmitting raw data in all ten cycles. This corresponds to an energy reduction of approximately 88.4%, demonstrating the substantial benefit of edge intelligence in reducing communication overhead and extending operational lifetime for battery-powered capsule systems.

### IV. CONCLUSION

This study presents a capsule-sized, multi-wavelength optical sensing system for order-of-magnitude–level classification of gastrointestinal bleeding flow rate, addressing the unmet clinical need for continuous, site-specific monitoring during the high-risk post-endoscopic period. By integrating controlled optical sensing, data-driven classification, and embedded deployment, the proposed framework extends capsule-based diagnostics beyond binary blood detection toward more informative assessment of bleeding severity.

Physics-based validation experiments confirmed that the measured wavelength-dependent transmission behavior is consistent with known hemoglobin absorption properties, establishing a reliable physical prior for subsequent learning-based analysis. Building upon this foundation, a lightweight 2D convolutional neural network achieved robust discrimination between non-bleeding interference and multiple bleeding flow-rate levels, achieving an overall classification accuracy of 98.75% under simulated gastric conditions.

To demonstrate practical feasibility, the proposed network was successfully deployed on a low-power microcontroller, where on-device inference incurred negligible computational energy. System-level power measurements further showed that performing local inference prior to wireless transmission can reduce overall energy consumption by approximately 88.4%, highlighting the advantage of edge intelligence for long-term, battery-powered operation.

While the current study focuses on controlled in vitro validation, future work will extend the proposed approach to in vivo evaluation and investigate more complex scenarios involving mixed blood–interference conditions and dynamically varying bleeding rates. Overall, the results demonstrate the feasibility of combining multi-spectral optical sensing and embedded intelligence to enable continuous, quantitative monitoring of gastrointestinal bleeding in realistic clinical settings.